\documentstyle[aps,epsf,multicol,psfig]{revtex}
\begin{document}
\draft
\title{Mesoscopic thermal transport through a weak link}
\author{Michael R. Geller and Kelly R. Patton}
\address{Department of Physics and Astronomy, University of Georgia, Athens, 
Georgia 30602-2451}
\date{July 8, 2001}
\maketitle

\begin{abstract}
We consider mesoscopic thermal transport between two bulk dielectrics joined 
by a narrow wire or weak mechanical link. In the ``tunneling'' regime where 
the phonon transmission probability through the link is small and the thermal 
conductance is much less than $\pi k_{\rm B}^2 T /6\hbar$, the thermal current
is determined by a product of the local vibrational spectral densities of the 
two bodies. We derive an expression for the thermal current that is a thermal 
analog of the well-known formula for the electrical current through a 
tunneling barrier.
\end{abstract}

\vskip 0.05in

\begin{multicols}{2}

\section{introduction}

Recently there has been considerable interest in the physics of phonons in
mesoscopic and nanoscale systems. In a beautiful experiment, Schwab {\it et 
al.} \cite{Schwab etal} observed a quantization of the thermal conductance in 
freely suspended one-dimensional dielectric wires, corresponding to a 
conductance given by $\pi k_{\rm B}^2 T /6\hbar$ per transmitted vibrational 
mode, analogous to the well-known electrical conductance quantization in units
of $e^2 /2 \pi \hbar$ per spin-resolved channel in one-dimensional mesoscopic 
conductors \cite{Beenakker review}. The thermal conductance quantization can 
be understood \cite{Rego and Kirczenow,Angelescu etal,Blencowe} by using a 
thermal analog of the Landauer-B\"uttiker formula \cite{Beenakker review}.

\begin{figure}
\centerline{\psfig{file=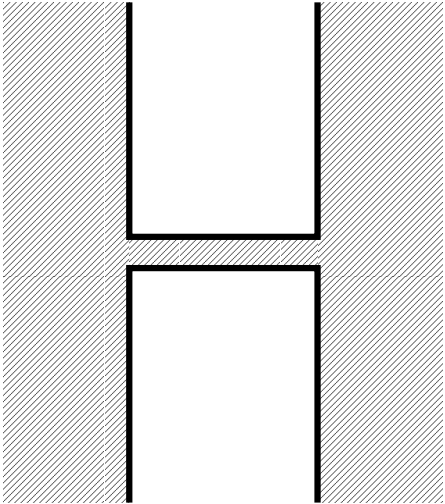,width=2.5in}}
\vspace{0.1in}\setlength{\columnwidth}{3.2in}
\centerline{\caption{Mesoscopic wire connected nonadiabatically to bulk 
dielectrics.
\label{wire figure}}}
\end{figure}

In this paper we consider mesoscopic thermal transport through a wire or weak 
link in the limit where the phonon transmission probability through the wire 
is small (the so-called nonadiabatic regime) and the thermal conductance is 
much less than $\pi k_{\rm B}^2 T /6\hbar$, the value corresponding to one 
fully propagating channel. The geometry we consider is shown schematically in 
Fig.~\ref{wire figure}.

The mechanical bridge may consist of one or more 
chemical bonds, or by a narrow wire of dielectric material, both of which we 
model by a harmonic spring of stiffness $K$ \cite{model footnote}. We obtain a general expression for the thermal current that can be regarded as a thermal 
analog of the well-known formula, derived by Schrieffer {\it et 
al.}\cite{Schrieffer etal,Mahan}, for the electrical current through a 
tunneling barrier. Our result can also be interpreted as an application of the
thermal Landauer formula in the weak-tunneling limit, with the 
energy-dependent phonon transmission probability calculated microscopically.

\section{thermal current in the weak-tunneling limit}

We consider two macroscopic dielectrics, L and R, held at fixed temperatures 
$T_{\rm L}$ and $T_{\rm R}$. The Hamiltonian of the isolated solids is $ H_0 
= H_{\rm L} + H_{\rm R},$ where
\begin{equation}
H_I \equiv \sum_n \omega_{In} \, a_{In}^\dagger \, a_{In}, \ \ \ \ \ \ \ 
I={\rm L,R}.
\label{unperturbed hamiltonian}
\end{equation}
The $a_{nI}^\dagger$ and $a_{nI}$ are phonon creation and annihilation 
operators for the left and right sides. The vibrational modes of the isolated 
bodies are labeled by $n$ and have energies $\omega_{In}$. 

\begin{figure}
\centerline{\psfig{file=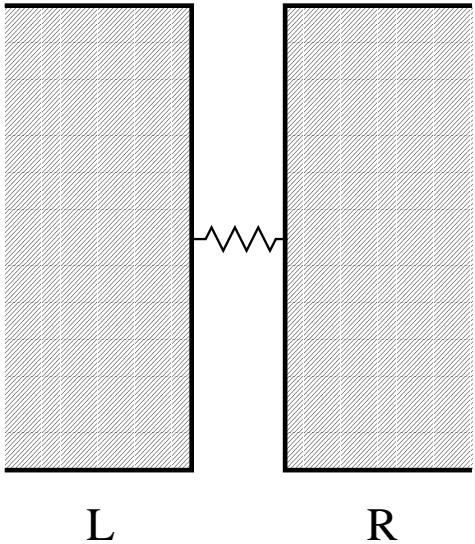,width=2.5in}}
\vspace{0.1in}\setlength{\columnwidth}{3.2in}
\centerline{\caption{Our weak link model. Two macroscopic dielectrics are 
joined by a harmonic spring.
\label{weak link figure}}}
\end{figure}

The two bodies are connected by a harmonic spring with stiffness $K$,
\begin{equation}
\delta H = {\textstyle{1 \over 2}} K \big(u_{\rm L}^z - u_{\rm R}^z \big)^2.
\label{interaction definition}
\end{equation}  
Here $u_{I}^z$ is the normal component of the displacement field ${\bf u}(
{\bf r})$ at the surface of body $I$ at the point of connection to the weak 
link, with the local surface normal taken to be in the $z$ direction. 
Our mesoscopic 
weak-link model is illustrated in Fig.~\ref{weak link figure}.

The system is described by the Hamiltonian $H = H_0 + \delta H$. The thermal
current is calculated by defining a thermal current operator 
${\hat I}_{\rm th}$ according to 
\begin{equation}
{\hat I}_{\rm th} \equiv \partial_t H_{\rm R} = i[H,H_{\rm R}].
\end{equation}
The expectation value of ${\hat I}_{\rm th}$ is 
the energy per unit time flowing from the left to the right body. To leading 
order in $\delta H$ (the weak-coupling limit) we 
obtain\cite{Patton and Geller thermal PRB}
\begin{equation}
I_{\rm th} = {2 \pi K^2 \over \hbar} \int_0^\infty \! d\epsilon \, \epsilon 
\, N_{\rm L}(\epsilon) \, N_{\rm R}(\epsilon) \, \big[ n_{\rm L}(\epsilon)
- n_{\rm R}(\epsilon) \big],
\label{thermal current}
\end{equation}
an expression analogous to the formula derived by Schrieffer {\it et al.} 
\cite{Schrieffer etal,Mahan} for the electrical current through a tunneling 
barrier. Here $n_{\rm L}(\epsilon)$ and $n_{\rm R}(\epsilon)$ are Bose 
distribution functions with temperatures $T_{\rm L}$ and $T_{\rm R}$, and 
\begin{equation}
N_I(\omega) \equiv - {\textstyle{1 \over \pi}} \, {\rm Im} \, D_I(\omega)
\label{spectral density definition}
\end{equation}
is the local spectral density, defined in terms of the Fourier transform of 
the retarded correlation functions
\begin{equation}
D_I(t) \equiv -i \theta(t) \big\langle [ u_I^z(t), u_I^z(0) ]\big\rangle_0
\label{propagator definition}
\end{equation}
for the isolated macroscopic bodies L and R. The spectral density defined in
(\ref{spectral density definition}) is different than the ordinary 
thermodynamic DOS, even in a homogeneous system. The quantity (\ref{spectral
density definition}), however, is the one relevant here.

Our result (\ref{thermal current}) shows that the (linear) thermal conductance
of a weak mesoscopic link, defined by 
\begin{equation}
G_{\rm th} \equiv \lim_{T_{\rm L} \rightarrow T_{\rm R}} \ {I_{\rm th} \over
T_{\rm L}-T_{\rm R}},
\end{equation}
will vary at low temperatures as a power-law in $T$. If we let $\alpha$ be the
exponent characterizing the power-law phonon spectral density of the 
reservoirs at low energies, then
\begin{equation}
G_{\rm th} \propto T^{2\alpha +1}.
\label{conductance power-law}
\end{equation}
For example, $\alpha = 1$ at the planar surface of a semi-infinite isotropic 
elastic continuum. 

A simple application of our theory to a nanostructure consisting of a 
cylindrical neck of Si material connecting two semi-infinite Si crystals has 
been given in Ref.~\cite{Patton and Geller thermal PRB}. 

\section{discussion}

The classical theory of thermal conduction, based on the heat equation and on 
the concept of a local thermal conductivity, is entirely inapplicable to 
mesoscopic dielectrics. In a mesoscopic system, thermal resistance is caused 
by elastic scattering of phonons, whereas in an infinite, disorder-free 
crystal it is caused by anharmonicity. In the non-mesoscopic regime (for 
example, at higher temperatures), the thermal conductivity of a narrow bridge 
is determined by the bridge material's bulk thermal conductivity $\kappa$, but
in the mesoscopic regime it is determined by the {\it mechanical} properties 
of the bridge material, through the stiffness $K$.

\acknowledgements

This work was supported by NSF CAREER Grant No.~DMR-0093217, and by a Research
Innovation Award and Cottrell Scholars Award from the Research Corporation. It
is a pleasure to thank Andrew Cleland, Steve Lewis, and Aleksandar Milosevic 
for useful discussions.

\end{multicols}



\begin{references}
\bibitem{model footnote} This model assumes that the forces between the two 
reservoir surfaces, transferred by the link, are mostly longitudinal.
\end{references}
\end{document}